# Enhancement of *in-field $J_c$* in MgB$_2$/Fe wire using single and multi-walled nanotubes

**J. H. Kim[a)], W. K. Yeoh, M. J. Qin, X. Xu, S. X. Dou**

*Institute for Superconducting and Electronic Materials, University of Wollongong, Northfields Avenue, Wollongong, New South Wales (NSW) 2522, Australia*

**P. Munroe**

*Electron Microscopy Unit, University of New South Wales, Kensington, Sydney, New South Wales (NSW) 2052, Australia*

**H. Kumakura, T. Nakane, C. H. Jiang**

*Superconducting Materials Center, National Institute for Materials Science, 1-2-1, Sengen, Tsukuba, Ibaraki 305-0047, Japan*

[a)] Electronic mail: jhk@uow.edu.au

# ABSTRACT

We investigated the doping effects of SWCNTs and MWCNTs on the $T_c$, lattice parameters, $J_c(B)$, microstructure, and $H_{c2}$ of MgB$_2$/Fe wire. These effects systematically showed the following sequence for $T_c$ and the $a$-axis: the SWCNT doped wire < the MW$_{short}$CNT doped wire < the MW$_{long}$CNT doped wire < un-doped wire, while $J_c(B)$ followed the sequence of the SWCNT doped wire > the MW$_{short}$CNT doped wire > the MW$_{long}$CNT doped wire > un-doped wire. A dominating mechanism behind all these findings is the level of C substitution for B in the lattice. The best $J_c(B)$ and $H_{c2}$ were obtained on SWCNT doped wire because the level of C substitution for B in this wire is higher than all others.

The effects of carbon (C) doping on superconducting properties in $MgB_2$ compound have been studied by a number of groups.[1-16] Early studies on C doping into $MgB_2$ have largely focused on the effect on superconductivity.[1-6] From the applications point of view, the effect of C doping on the flux pinning properties and upper critical field ($H_{c2}$) is crucially important. Recently, several groups have reported a significant improvement in critical current density ($J_c$), $H_{c2}$, and irreversibility field ($H_{irr}$) in $MgB_2$ through C doping in various forms, including nano-C, nano-SiC, carbon nanotube (CNT), and $B_4C$.[7-19] Among various carbon precursors, CNT are particularly interesting as their special geometry (high aspect ratio and nanometer diameter) may induce more effective pinning centers compared to other C-containing precursors. Furthermore, CNTs have unusual electrical, mechanical, and thermal properties.[20-23] These properties could improve the interior thermal stability, heat dissipation, and mechanical strength of $MgB_2$ superconductor wire. The authors' group has demonstrated that CNT doping not only resulted in a significant enhancement of *in-field* $J_c$ performance[9] but also improved heat transfer and dissipation[17,18]. The CNTs are composed of one or more concentric graphene cylinders, which are called single walled nanotubes (SWCNTs) and multi walled nanotubes (MWCNTs), respectively.[24] The fundamental properties of SWCNTs and MWCNTs are different from each other. The effect of different type of CNTs on the superconducting properties of $MgB_2$ remains unclear. Thus, it is necessary to study the effects of the configuration and dimensions of the CNTs on the superconducting properties of $MgB_2$.

$MgB_2$/Fe wires were prepared by *in-situ* reaction and the powder-in-tube method. Powders of magnesium (Mg, 99%), boron (B, 99%), and SWCNT/MWCNTs were used as starting materials.

These powders were well mixed with a starting composition of $MgB_{1.8}C_{0.2}$, because this composition resulted in the highest $J_c$ in our group of samples.[9] Two different MWCNTs batches had two different average aspect ratios and are hereafter called $MW_{short}CNT$ and $MW_{long}CNT$, respectively, referring to the length of the CNTs. The specifications of the CNTs are listed in Table I. The mixed powders were packed into iron (Fe) tubes, and then the composites were drawn to an outer diameter of 1.42 mm. These wires were then sintered at 800 to 1000°C for 30 minutes under high purity argon. The heating rate was 5°C/min. Un-doped $MgB_2$/Fe wire was also fabricated for reference and comparison by applying the same process. All samples were characterized by critical current ($I_c$), critical temperature ($T_c$), $H_{c2}$, X-ray diffraction (XRD), and transmission electron microscopy (TEM). The transport $I_c$ measurement was measured by the standard four-probe method at 4.2 K with criterion of 1 $\mu Vcm^{-1}$. $T_c$ was defined as the onset temperature at which diamagnetic properties were observed. In addition, $H_{c2}$ was defined as $H_{c2}=0.9R(T_c)$ from the resistance ($R$) vs. temperature ($T$) curve. A PW1730 X-ray diffractometer with Cu Kα radiation was used to determine the phase and crystal structure of all the samples. The lattice parameters were obtained from Rietveld refinement.

Figure 1 shows the $T_c$ for all CNT doped and un-doped $MgB_2$/Fe wires as a function of sintering temperature. It should be noted that for $MW_{short}CNT$ and SWCNT doped wires, $T_c$ decreased systematically as the sintering temperature increased, while $T_c$ of un-doped $MgB_2$/Fe wires increased with increasing sintering temperature. It is well established that higher sintering temperature results in better crystallinity, and hence higher $T_c$. On the other hand, for CNT doped samples, $T_c$ is suppressed in proportion to the amount of C substituted in a given sample.[9] Even though the nominal composition remains the same, a higher sintering temperature results in more C substitution for B. Thus, $T_c$ for the MW$_{short}$CNT and SWCNT doped wires decreased with increasing sintering temperature. The SWCNT doping showed a stronger depression in $T_c$ than the MW$_{short}$CNT doping, suggesting that SWCNT is more reactive with MgB$_2$ than MW$_{short}$CNT. What is surprising is that the $T_c$ for the MW$_{long}$CNT doped wire showed an opposite trend from the MW$_{short}$CNT and SWCNT doped wires, but had the same behavior as un-doped MgB$_2$/Fe wires. These results indicate that the MW$_{long}$CNTs are the least reactive while the MW$_{short}$CNTs are more reactive, and the SWCNTs are the most reactive to MgB$_2$. For the MW$_{long}$CNT doped wire, there are two conflicting factors that affect $T_c$. The increase in sintering temperature improves both crystallinity and C substitution for B, the former will increase $T_c$ while the latter will decrease $T_c$. Because of the low reactivity of MW$_{long}$CNTs the former factor dominates, the $T_c$ increases with sintering temperature as shown in Figure 1.

These observations are further supported by the XRD data as shown in Table II. Within the limits of calculation error, the $a$ axis lattice parameter for MW$_{long}$CNT doped MgB$_2$/Fe wires showed little change at the sintering temperature of 900°C compared with un-doped one. In contrast, the $a$ axis parameters for MW$_{short}$CNT and SWCNT doped wires showed a noticeable decrease compared with the un-doped and MW$_{long}$CNT doped ones. The $c$-axis lattice parameter remained unchanged for all doped samples. This is the typical situation for C substitution in B site as reported previously.[2,6] Based on the lattice parameter changes[25] we could quantitatively estimate the amount of C substituted in B sites as shown in Table II. It is evident that the amount of C substitution for B is much less than the nominal composition. However, what is worth noting is that more C is substituted in B sites in the SWCNT and MW$_{short}$CNT doped wires than the MW$_{long}$CNT doped ones. As a result, we believe that the differences in the actual substitution with C for the different CNTs is most likely due to the different reactivity, which is consistent with the results of $T_c$ depression, as shown in Figure 1.

Figure 2 shows the magnetic field dependence of transport $J_c$ of all CNT doped and un-doped MgB$_2$/Fe wires at 4.2 K. It was found that the $J_c$ values are spread out by far more than an order of magnitude in the field region measured. For example, the $J_c$ values at 4.2 K and 12 T for the SWCNT doped wire are higher than that of the un-doped wire by a factor of 35 when all were sintered at 900°C. Of particular interest is that the trend of $J_c(B)$ followed the same sequence of change as for $T_c$ and the $a$-axis. That is, the SWCNT doped wires showed the best performance in $J_c(B)$, and the MW$_{short}$CNT doped wires were next, while the MW$_{long}$CNT doped ones had the least improvement in $J_c(B)$ compared to the un-doped ones. What is more interesting is that in relation to the sintering temperature, the $J_c(B)$ for both the MW$_{short}$CNT and SWCNT doped wires showed an increase with increasing sintering temperature while the MW$_{long}$CNT doped and un-doped wires followed the opposite trend. It has been reported that C substitution in B sites can improve the $J_c(B)$, $H_{irr}$, and $H_{c2}$, but depresses $T_c$ for MgB$_2$.[7-19] The $J_c(B)$ of un-doped MgB$_2$/Fe wire decreased with increasing sintering temperature as a result of further improvement of crystallinity at higher sintering temperature.[26] The MW$_{long}$CNT doped wires showed the same trend as the un-doped ones, suggesting that the level of C substitution for B did not increase much with increasing sintering temperature. Thus, the improvement of crystallinity dominated the $J_c(B)$ behaviour in the same way as for as un-doped wires.

It is well established that C substitution into B sites results in an enhancement in $J_c(B)$ and $H_{irr}$[27]. The temperature dependence of the normalized $H_{c2}$ for all CNT doped and un-doped MgB$_2$/Fe wires sintered at 900°C is shown in Figure 3. The 10wt%SiC doped MgB$_2$/Fe were also included for comparison and reference.[28] The best $H_{c2}$ was obtained on a SWCNT doped sample, because the slope of d$H_{c2}$/d$T$ for the SWCNT sample is larger compared with those for both MWCNTs. This is because SWCNT contributed more C to B sites at high sintering temperature, which is believed to increase the intra-band scattering, as well as shorten the mean free path and coherence length.[26,29]

From the $R$ vs. $T$ curve (not shown), we also calculated the resistivity of the un-doped, MW$_{short}$CNT, MW$_{long}$CNT, and SWCNT doped samples sintered at 900°C as shown in table II. It is clear that the SWCNT doped samples showed a relatively higher value of resistivity than the MW$_{short}$CNT doped ones while the un-doped one had a lower value of resistivity. The increased resistivity for the SWCNT doped sample could be due to the increased impurity scattering, as it has been shown that substitution of C for B can result in strong σ scattering.[30] The increased resistivity of SWCNT doped MgB$_2$ due to the substitution of C for B would contribute to the enhancement in $J_c(B)$ and $H_{c2}$ in the SWCNT doped sample. However, the MW$_{long}$CNT doped sample had a lower resistivity, compared to the pure sample.

Figure 4 shows TEM images for (a) MW$_{short}$CNT, (b) SWCNT, and (c) MW$_{long}$CNT doped

MgB$_2$/Fe wires sintered at 900$^o$C. The CNTs still existed as MgB$_2$-CNT composites after sintering, in particular, for MW$_{long}$CNT doped sample. These CNTs can contribute to improvement of thermal and mechanical properties[17,18]. TEM images showed that most CNTs have at least one open-end. It is noted that MW$_{long}$CNTs are well intact and clearly visible after heat treatment (Figure 4(c)) compared to SWCNTs and MW$_{short}$CNTs. Together with the results on $T_c$, $J_c(B)$, and the lattice parameters it is reasonable to believe that the difference in reactivity of CNTs is attributable to the number of open-ends of CNTs. As the SWCNTs have a much smaller diameter (only 1-2 nm) compared to MWCNTs the number of open-ends in SWCNT doped wire is larger than those in MW$_{short}$CNT doped wire. As MW$_{long}$CNTs are much longer than SWCNTs and MW$_{short}$CNTs the number of open-ends is much smaller than for the latter two. Thus, there is little substitution reaction between MW$_{long}$CNTs and MgB$_2$.

In summary, SWCNT is an attractive dopant for enhancing $J_c$ of MgB$_2$ superconductor in the high field region. The $J_c$ in 12 T and 4.2 K for the SWCNT doped wire sintered at 900$^o$C increased by a factor of 35 compared to that of the un-doped wire. The observed $J_c(B)$ enhancement in the SWCNT doped sample is attributable to the high level of C substitution into B sites. This demonstrates that C substitution for B from dopants is essential for enhancement of $J_c(B)$ and $H_{c2}$. Doping with MW$_{long}$CNT has very level of C substitution for B the improvement in $J_c$ and $H_{c2}$ is insignificant.

The authors thank Dr. T. Silver, Dr. J. Horvat, and R. Kinnell for their helpful discussions. This work was supported by the Australian Research Council, Hyper Tech Research Inc., USA, Alphatech International Ltd., NZ, and the University of Wollongong.

[1] T. Takenobu et al., Phys. Rev. B **64**, 134513 (2001).

[2] W. Mickelson et al., Phys. Rev. B **65**, 052505 (2002).

[3] Z. H. Cheng et al., J. Appl. Phys. **91**, 7125 (2002).

[4] R. A. Ribeiro et al., Physica C **382**, 166 (2002).

[5] J. Wei, Y. Li et al.. Chem. Phys. **78**, 785 (2003).

[6] E. Ohmichi et al., J. Phys. Soc. Jpn. **73**, 2065 (2004).

[7] S. X. Dou et al., Appl. Phys Lett. **81**, 3419 (2002).

[8] S. Soltanian et al., Physica C **390**, 185 (2003).

[9] S. X. Dou et al., Appl. Phys. Lett. **83**, 4996 (2003).

[10] B. J. Senkowicz et al., Appl. Phys. Lett. **86**, 202502 (2005).

[11] R. H. T. Wilke et al., Phys. Rev. Lett. **92**, 217003 (2004).

[12] V. Braccini et al., Phys. Rev. B **71**, 012504-1 (2004).

[13] H. Kumakura et al., Appl. Phys. Lett. **84**, 3669 (2004).

[14] S. X. Dou et al., J. Appl. Phys. **96**, 7549 (2004).

[15] M. D. Sumption et al., Appl. Phys. Lett. **86**, 092507-1 (2005).

[16] A. Yamamoto et al., Supercond. Sci. Technol. **18**, 1323 (2005).

[17] S. X. Dou et al., Adv. Mater. **18**, 785 (2006).

[18] S. K. Chen et al., Appl. Phys. Lett. **87**, 182504 (2005).

[19] Y. Ma et al., Appl. Phys. Lett. **88,** 072502 (2006).

[20] R. H. Baughman et al., Science **297**, 787 (2002).

[21] B. Q. Wei et al., Appl. Phys. Lett. **79**, 1172 (2001).

[22] P. Kim et al., Phys. Rev. Lett. **87**, 215502 (2001).

[23] M. M. J. Treacy et al., Nature **381**, 678 (1996).

[24] E. Thostenson et al., Comp. Sci. Technol. **61**, 1899 (2001).

[25] S. Lee et al., Physica C **397**, 7 (2003).

[26] J. H. Kim et al., J. Appl. Phys. **100**, 013908 (2006).

[27] A. Yamamoto et al., Appl. Phys. Lett **86**, 212502 (2005).

[29] S. Soltanian et al., Supercond. Sci. Technol. **18**, 658 (2005).

[30] I. I. Mazin et al., Phys. Rev. Lett. **89**, 1070021 (2002).

**Table I.** The specifications of $MW_{short}CNT$, $MW_{long}CNT$, and SWCNT.

|  | $MW_{short}CNT$ | $MW_{long}CNT$ | SWCNT |
| --- | --- | --- | --- |
| Purity | >95% | >95% | >90% |
| Outer diameter (nm) | 20-30 | <8 | 1-2 |
| Length (μm) | 0.5 | 0.5-200 | 5-15 |
| Impurity components | Cl, Fe, Ni | Al, Cl, Co, S | amorphous C, Mg, Co, Mo, $SiO_2$ |

**Table II.** Lattice parameters, actual C substitution[25] (extrapolation from measured lattice parameters), and resistivities of all CNT doped and un-doped $MgB_2$/Fe wires sintered at 900°C for 30 min with a starting composition of $MgB_{1.8}C_{0.2}$.

| Samples | Lattice parameters | | | Actual C substitution (x) in $MgB_{2-x}C_x$ | $\rho_{40K}$ (μΩcm) |
|---|---|---|---|---|---|
| | $a$ (Å) | $c$ (Å) | $c/a$ | | |
| Un-doped | 3.082 | 3.524 | 1.1434 | | 24.8 |
| $MW_{short}CNT$ | 3.073 | 3.525 | 1.1468 | 0.041 | 57.5 |
| $MW_{long}CNT$ | 3.078 | 3.524 | 1.1449 | 0.018 | 2.60 |
| SWCNT | 3.071 | 3.524 | 1.1475 | 0.050 | 69.7 |

**FIGURE CAPTIONS**

**FIG. 1.** $T_c$ for all CNT doped and un-doped $MgB_2$/Fe wires as a function of sintering temperature. $T_c$ was defined as the onset temperature at which diamagnetic properties were observed.

**FIG. 2.** $J_c(B)$ for MWCNT and SWCNT doped $MgB_2$/Fe wires sintered at various temperatures for 30 min. A $J_c(B)$ curve of an un-doped $MgB_2$ wire is also shown for comparison and reference.

**FIG. 3.** Temperature dependence of normalized $H_{c2}$ for all CNT doped and un-doped $MgB_2$/Fe wires sintered at 900°C for 30 min. The $H_{c2}$ for 10wt% SiC doped $MgB_2$ samples are also shown for comparison.[29]

**FIG. 4.** TEM images for (a) $MW_{short}CNT$, (b) SWCNT, and (c) $MW_{long}CNT$ doped $MgB_2$/Fe wires sintered at 900°C for 30 min with the nominal composition of $MgB_{1.8}C_{0.2}$.

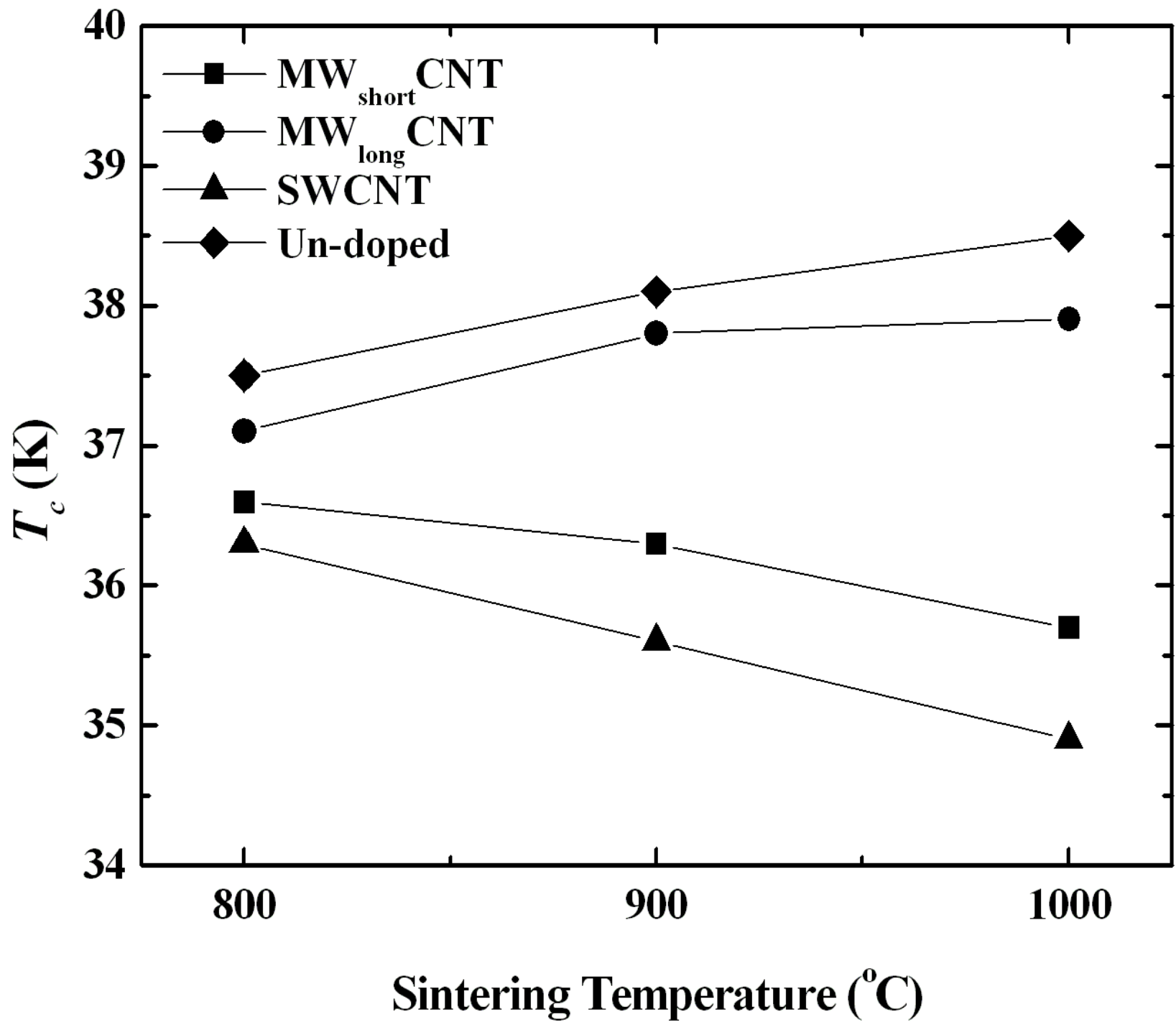

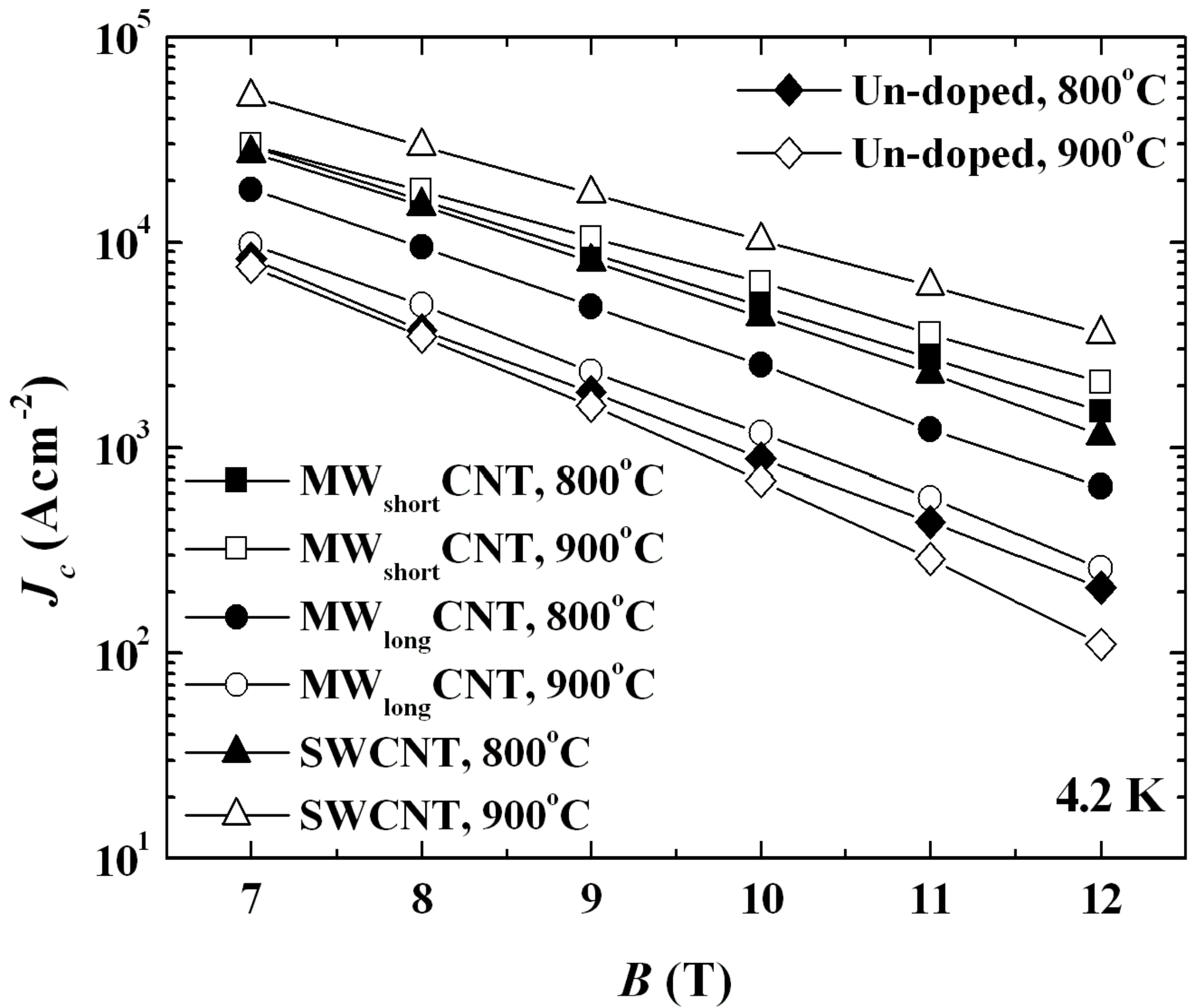

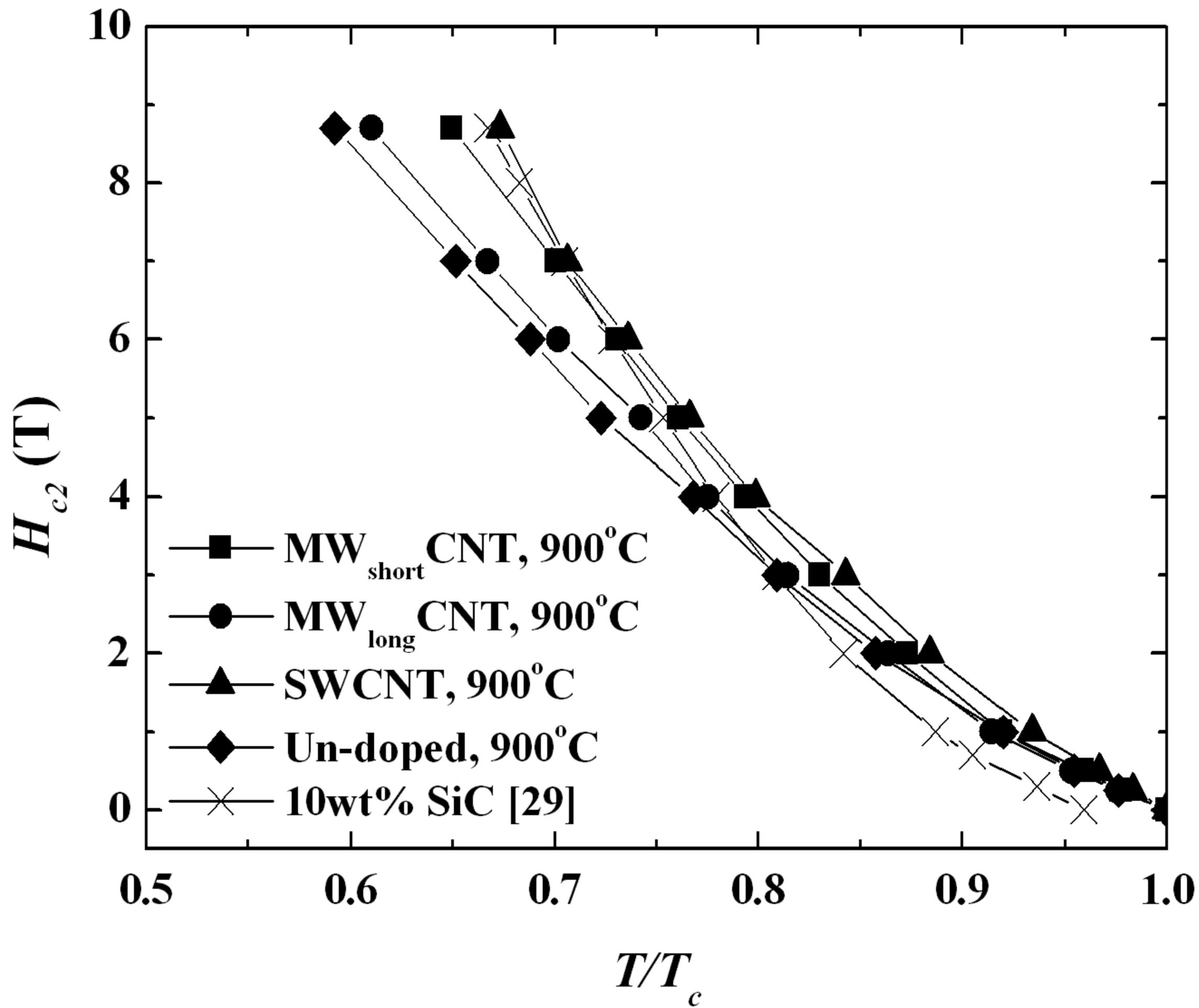

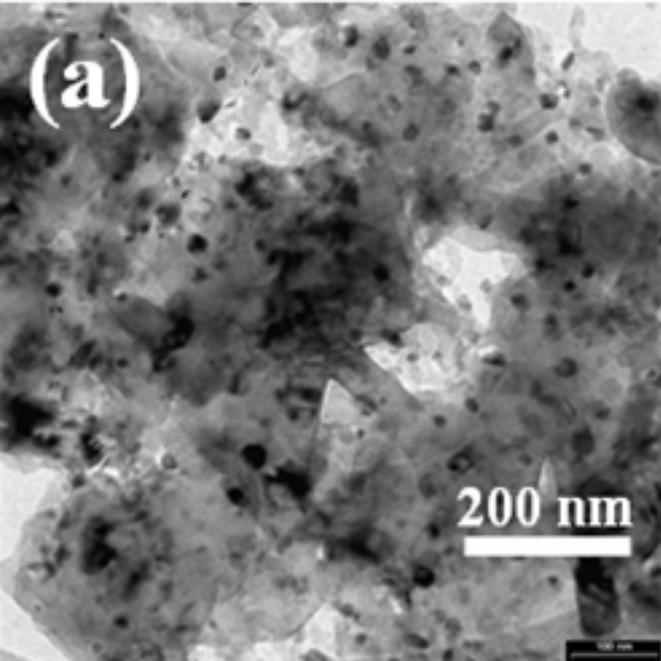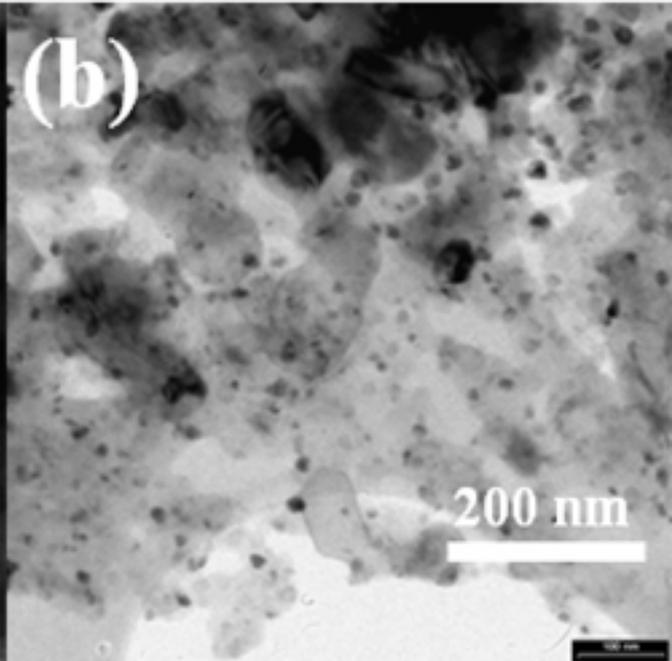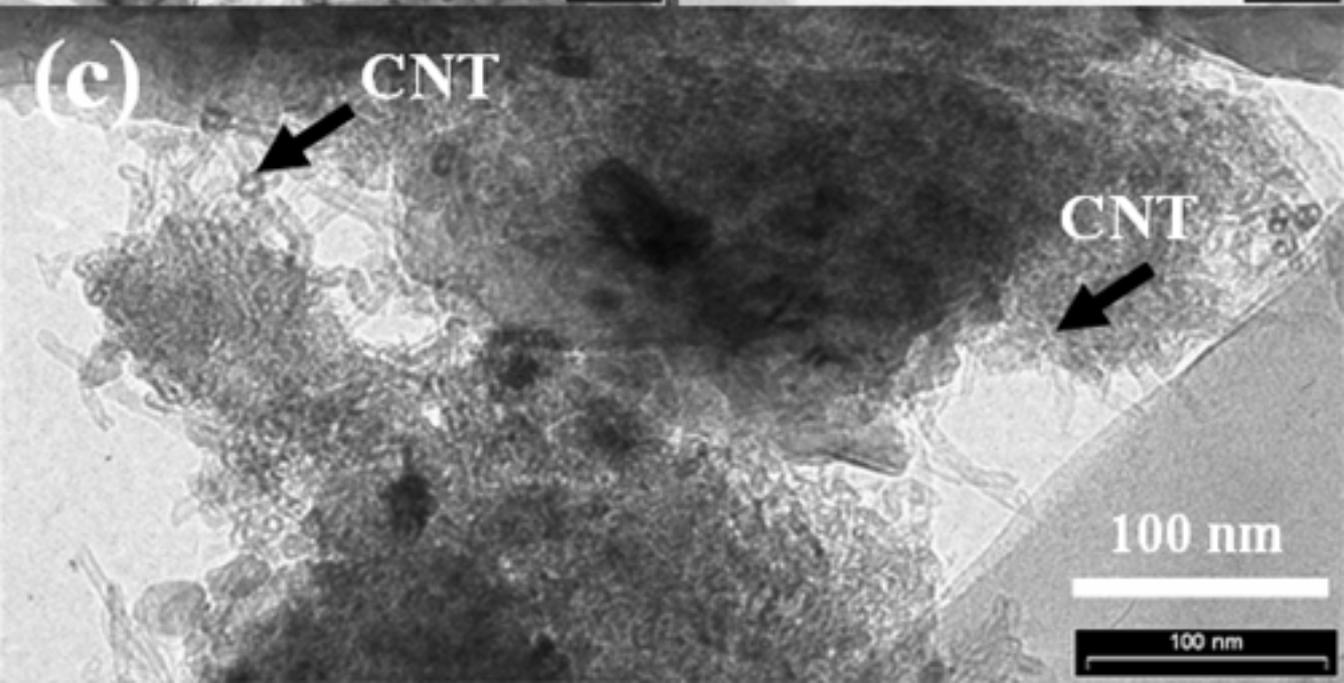